\documentclass[%
    reprint,
    amsmath,
    amssymb,
    aps,
    prstab,
]{revtex4-2}

\usepackage[export]{adjustbox}
\usepackage{graphicx} 
\usepackage{dcolumn}
\usepackage{bm}
\usepackage[caption=false]{subfig}
\usepackage{appendix}
\usepackage[hidelinks]{hyperref}

\newcommand{\vect}[1]{\boldsymbol{\mathbf{#1}}}

\begin{document}

\title{Four-dimensional phase space tomography from one-dimensional measurements of a hadron beam}

\author{Austin Hoover}%
\email{hooveram@ornl.gov}
\affiliation{Oak Ridge National Laboratory, Oak Ridge, Tennessee 37830, USA}

\date{\today}

\begin{abstract}

In this paper, we use one-dimensional measurements to infer the four-dimensional phase space density of an accumulated proton beam in the Spallation Neutron Source (SNS) accelerator. The reconstruction was performed by maximizing the distribution's entropy subject to the measurement constraints, and thus represents the most conservative inference from the data. The reconstructed distribution reproduces the measured profiles down to the noise level, and simulations indicate that the problem is reasonably well-constrained. Similar measurements could serve as benchmarks for beam dynamics simulations in the SNS or hadron accelerators.
\end{abstract}

\maketitle

\section{Introduction}\label{sec:introduction}

High-dimensional phase space measurements provide valuable information about the beam dynamics in an accelerator; however, conventional diagnostics can only measure the distribution's low-dimensional projections. Reconstructing the phase space distribution from these projections is referred to as \textit{phase space tomography}.

Let us focus on the specific case of four-dimensional phase space. Recent experiments have demonstrated four-dimensional tomography from two-dimensional measurements in electron accelerators \cite{Wolski_2020, Wolski_2023, Roussel_2023}, as well as in an H- linac using laser wires \cite{Wong_2022_ment}. However, two-dimensional measurements are not always available. After the initial acceleration stages in high-power accelerators, the primary diagnostics are \textit{wire scanners}, which record the beam density on a one-dimensional axis. It is not \textit{a priori} obvious whether a realistic number of one-dimensional views can tightly constrain a four-dimensional distribution, nor whether the required views can be obtained in an accelerator.

To our knowledge, the only published 4D-from-1D reconstruction was by Minerbo, Sander, and Jameson in 1981 \cite{Minerbo_1981}. The authors fit a four-dimensional distribution to 10 profiles. The sole figure in the paper compares the reconstructed $x$-$x'$ and $y$-$y'$ distributions to direct measurements. This comparison says nothing about the accuracy in the four-dimensional space; further study is needed to determine the usefulness of four-dimensional tomography when only one-dimensional measurements are available.

To this end, in this paper, we use one-dimensional measurements to infer the four-dimensional phase space distribution of an accumulated proton beam in the Spallation Neutron Source (SNS) accelerator. We measure the beam using four wire scanners in a transfer line between the proton accumulator ring and the spallation target. Following \cite{Minerbo_1981}, we reconstruct the distribution using the method of maximum relative entropy. The reconstructed distribution reproduces the measured profiles down to the wire scanner noise level.

As a form of uncertainty quantification, we simulate the reconstruction with fake data generated by various ground truth distributions. We find that the two-dimensional marginal distributions ($x$-$x'$, $y$-$y'$, $x$-$y$, $x$-$y'$, $y$-$x'$, $x'$-$y'$) are reasonably well constrained by the data. We also find that these low-dimensional views can obscure certain higher-dimensional features, such as four-dimensional hollowing. We conclude that our set of optics is sufficient for typical beams in the SNS, but that different optics or additional measurements may be needed to reconstruct arbitrary distributions in the full four-dimensional phase space.

\textit{Organization}: In Section~\ref{sec:reconstruction_algorithm}, we review the method of maximum relative entropy. In Section~\ref{sec:methods}, we describe the SNS accelerator, diagnostics, and experimental details. In Section~\ref{sec:results}, we report the reconstruction results and perform a suite of reconstructions from fake data. Section~\ref{sec:conclusion} concludes.

\section{Reconstruction algorithm}\label{sec:reconstruction_algorithm}

We use the method of maximum relative entropy to infer the phase space distribution from its one-dimensional projections. Denote the distribution function as $f(\vect{v})$, where $\vect{v} = (v_1, \dots , v_N)$ is the phase space coordinate vector. (In four-dimensional phase space, $\vect{v} = (x, x', y, y')$, where $x$ and $y$ are the positions and $x'$ and $y'$ are the momenta.) We assume the $i$th projection is measured after a linear transformation
\begin{equation}\label{eq:forward}
    \vect{w}_i = \vect{M}_i \vect{v},
\end{equation}
where $\vect{M}_i$ is an $N \times N$ matrix representing the accelerator transport between the reconstruction point and the measurement device, and $\mathbf{w}_i = (w_{i1}, \dots, w_{iN})$ is the transformed coordinate vector. We measure the projected density $g_i(w_{ij})$. Thus, the distribution must satisfy the following constraints:
\begin{equation}\label{eq:constraints}
    g_i(w_{ij}) - \int 
    f(\vect{v}(\vect{w}_i))
    \prod_{k \ne j} dw_{ik} = 0
\end{equation}

To select a distribution compatible with Eq.~\eqref{eq:constraints}, we update a \textit{prior} distribution $f_*(\vect{v})$ to a posterior by maximizing the relative entropy
\begin{equation}
    S[f(\vect{v}), f_*(\vect{v})]
    = 
    -
    \int f(\vect{v})
    \log
    \left(
        \frac{f(\vect{v})}{f_*(\vect{v})}
    \right) 
    d\vect{v},
\end{equation}
subject to Eq.~\eqref{eq:constraints}. The method of Lagrange multipliers leads to the following form of the posterior:
\begin{equation}\label{eq:ment:posterior}
    f(\vect{v}) = f_*(\vect{v}) \prod_{i} \exp ( \lambda_i(w_{ij}) ),
\end{equation}
where $\lambda_i(w_{ij})$ are Lagrange multiplier functions. The MENT algorithm uses a nonlinear Gauss-Seidel relaxation method to find $\lambda_i$ such that the distribution in Eq.~\eqref{eq:ment:posterior} satisfies the constraints in Eq.~\eqref{eq:constraints}. We use the particle-based version of MENT described in \cite{Hoover_2024_ment}.

An advantage of the maximum entropy method is the ability to incorporate external information through the prior $f_*(\vect{v})$.  Following \cite{Minerbo_1981}, because the covariance matrix  ${\vect{\Sigma}} = \langle \vect{v}\vect{v}^T\rangle$ is \textit{overdetermined} by the measured second-order moments, we consider $\vect{\Sigma}$ to be prior information. For best-fit covariance matrix $\hat{\vect{\Sigma}}$, we define the following Gaussian prior:
\begin{equation}\label{eq:ment_prior}
    f_*(\vect{v}) = 
    \left[
        \frac{1}{(2\pi)^{N} |\vect{\hat\Sigma}|}
    \right]^{1/2}
    \exp \left(
        -\frac{1}{2} \vect{v}^T \hat{\vect{\Sigma}}^{-1} \vect{v}
    \right),
\end{equation}
(The Gaussian distribution maximizes the absolute entropy subject to a known covariance matrix.) There is no requirement to define the prior in this way, but it is a reasonable choice in accelerator physics when the covariance matrix is known.

We use the following standard recipe to fit the covariance matrix to the measured second-order moments. Under the symplectic linear transformation $\vect{M}_i$, $\vect{\Sigma}$ transforms as
\begin{equation}\label{eq:cov-transform}
    \mathbf{\Sigma} \rightarrow \mathbf{M}_i \mathbf{\Sigma} \mathbf{M}_i^T. 
\end{equation}
Eq.~\eqref{eq:cov-transform} linearly relates the measured second-order moments  $\eta_i = \langle w_{ij}^2 \rangle$ to the lower-triangular elements $\vect{\sigma}$ of the initial covariance matrix. Stacking the measurements into a single vector $\vect{\eta} = (\eta_1, \eta_2, \dots)$ gives the following system of equations:
\begin{align}\label{eq:llsq}
    \vect{A} \vect{\sigma} = \vect{\eta},
\end{align}
where $\vect{A}$ is determined by the transfer matrix elements. We generate a solution $\hat{\vect{\sigma}}$ (and therefore $\hat{\vect{\Sigma}}$) to Eq.~\eqref{eq:llsq} using ordinary linear least squares (LLSQ) \cite{Prat_2014}:
\begin{equation}
    \hat{\vect{\sigma}} = (\vect{A}^T\vect{A})^{-1} \vect{A}^T \vect{\eta}.
\end{equation}

After computing the LLSQ solution and resulting prior (Eq.~\eqref{eq:ment_prior}), we run the MENT algorithm to reconstruct the beam distribution in normalized coordinates,
\begin{equation}\label{eq:norm}
    \vect{z} = \vect{T}^{-1} \vect{v},
\end{equation}
such that $\langle \vect{z}\vect{z}^T \rangle = \vect{I}$. ($\vect{T}$ is a symplectic matrix which diagonalizes the covariance matrix: $\vect{\Sigma} = \vect{T}\vect{T}^T$.) When reconstructing in normalized two-dimensional phase space, the projection angles in the $x$-$x'$ plane are equivalent to the phase advances; this can be advantageous because the phase advances may be more uniformly spaced than the projection angles \cite{Hock_2013_ment}. Eq.~\eqref{eq:norm} generalizes this approach to higher dimensions. To reconstruct in the normalized coordinates, we append $\vect{T}$ to each transfer matrix:

\begin{equation}
    \vect{M}_i \rightarrow \vect{M}_i \vect{T},
\end{equation}
so that the matrix unnormalizes the coordinates before mapping them to the $i$th measurement location.

\section{Methods}\label{sec:methods}

\subsection{Accelerator and diagnostics}

\begin{figure*}
    \centering
    \includegraphics[width=\textwidth]{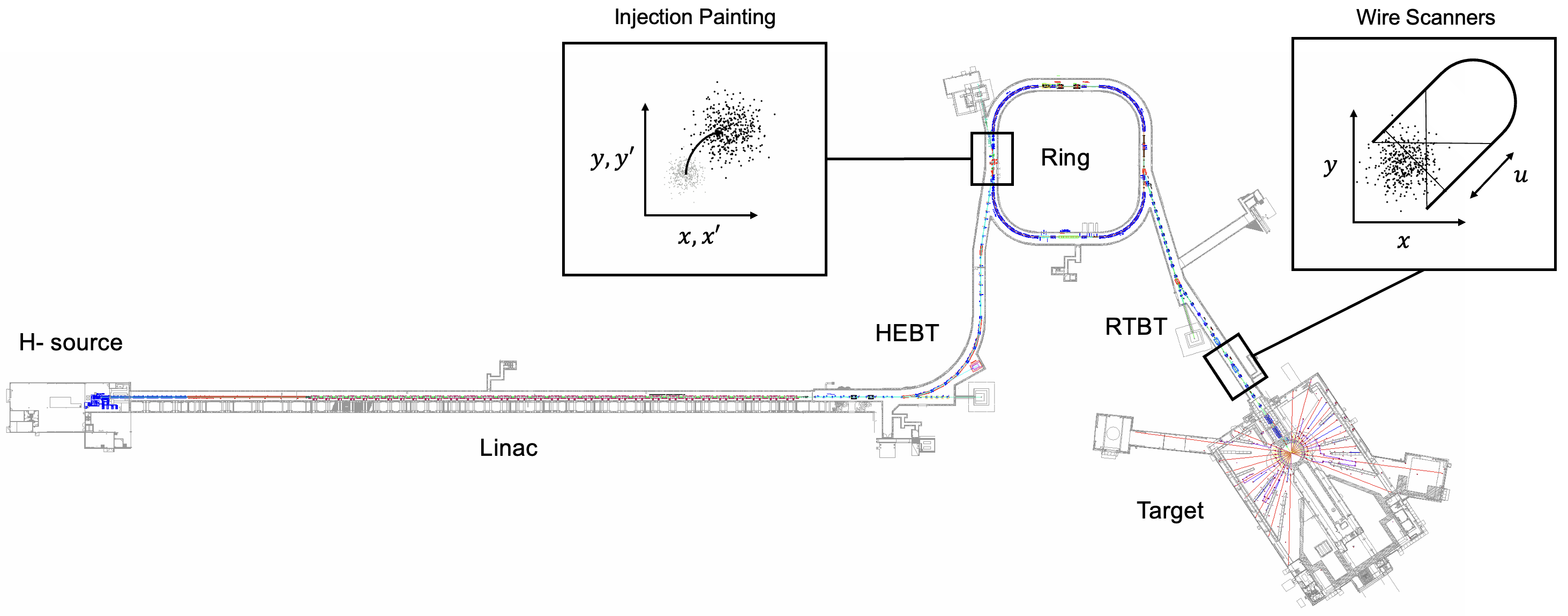}
    \caption{Layout of the Spallation Neutron Source (SNS) accelerator. H- ion \textit{minipulses} are repeatedly accelerated to 1 GeV in the linac and converted to protons by passing through a carbon foil at the ring injection point. A 700-nanosecond proton pulse is accumulated over 1000 turns in the ring. Dipole magnets vary the mean phase space coordinates ($x$, $x'$, $y$, $y'$) of the circulating beam relative to the injected beam as a function of time during accumulation. In the Ring-Target Beam Transport (RTBT), four wire scanners measure the particle density on a horizontal, vertical, and diagonal axis ($x$, $y$, $u$).}
    \label{fig:sns}
\end{figure*}

Our experiments were conducted at the Spallation Neutron Source (SNS) at Oak Ridge National Laboratory (ORNL), shown in Fig.~\ref{fig:sns}. The SNS bombards a liquid mercury target with intense proton pulses at 60 Hz repetition rate \cite{Henderson_2014}. Each pulse is the superposition of 1000 ``minipulses'' injected from a 400-meter linac into a 250-meter storage ring at 1 GeV kinetic energy. The final pulse contains approximately $1.4 \times 10^{14}$ protons, corresponding to 1.4 MW beam power \footnote{These numbers correspond to the date of our experiments. The current beam power at the SNS is 1.7 MW and will soon increase to 2.4 MW.}. The accumulated beam is extracted and transported from the ring to the target in a 150-meter Ring-Target Beam Transport (RTBT).

The diagnostics in the ring are limited to beam position monitors (BPMs) and beam loss monitors (BLMs). There are additional diagnostics in the RTBT; we focus on four wire scanners just before the target.
Each wire scanner is a set of three 100-$\mu$\textit{m} tungsten wires mounted on a fork. The secondary electron emission from the wires gives the beam density on the horizontal ($x$), vertical ($y$), and diagonal ($u$) axis in the transverse plane. Each wire scanner operates at 1 Hz repetition rate; thus, each point in the measured profile corresponds to a different beam pulse. A single measurement takes approximately five minutes, including the return to the initial wire position. The four wire scanners run in parallel, so each scan generates twelve profiles. Repeated measurements typically generate the same profile to within a few percent error \cite{Hoover_2022}.

\begin{figure}
    \centering
    \includegraphics[width=\columnwidth]{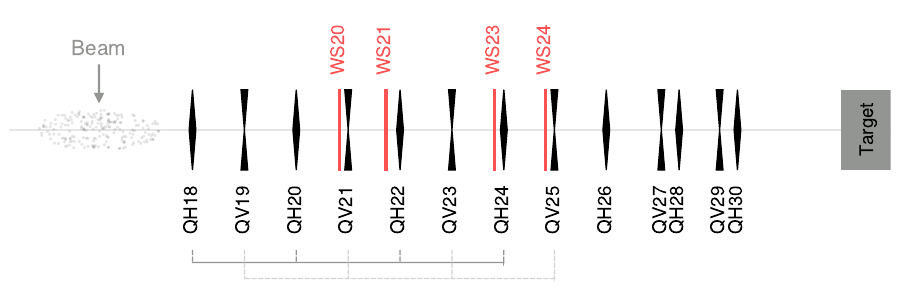}
    \caption{Wire scanner region of the RTBT. Each wire scanner (WS20, WS21, WS23, WS24) contains a horizontal, vertical, and diagonal wire. The faint lines at the bottom of the figure indicate quadrupoles with shared power supplies.}
    \label{fig:ws}
\end{figure}

Fig.~\ref{fig:ws} shows the locations of the wire scanners and quadrupole magnets. The first eight quadrupoles are controlled by two power supplies, while the remaining quadrupoles are controlled by independent power supplies. We aim to measure the particle distribution before the first quadrupole (QH18).

\subsection{Optics}

Given a finite number of measurements, it is not entirely clear how to determine the optimal set of views---those that place the tightest constraints on the unknown four-dimensional distribution. Nor is it entirely clear \textit{how many} views are required to reconstruct the distribution with sufficient accuracy. These questions deserve further study, especially when the focusing system has coupled or nonlinear elements.

Previous experiments have reconstructed four-dimensional distributions from two-dimensional $x$-$y$ projections in uncoupled focusing systems. Our focusing system is also uncoupled, and each wire scanner provides three one-dimensional projections that constrain the density in the $x$-$y$ plane. Therefore, we hypothesize that these experiments are relevant to our case. 

Hock and Wolski \cite{Hock_2013} discovered a nearly closed-form solution for the four-dimensional phase space density from a set of two-dimensional measurements. The solution assumes independent rotations in the $x$-$x'$ and $y$-$y'$ planes by the phase advances $\mu_x$ and $\mu_y$. In other words, in a normalized frame, the transfer matrices take the form
\begin{equation}
    \vect{M}_i =
    \begin{pmatrix}
        \phantom{-}\cos\mu_{x_i} & \sin\mu_{x_i} & 0 & 0 \\
        -\sin\mu_{x_i} & \cos\mu_{x_i} & 0 & 0 \\
        0 & 0 & \phantom{-}\cos\mu_{y_i} & \sin\mu_{y_i} \\
        0 & 0 & -\sin\mu_{y_i} & \cos\mu_{y_i}
    \end{pmatrix}.
\end{equation}
Scanning the phase advances over the entire $\mu_x$-$\mu_y$ plane in a nested loop reduces the four-dimensional reconstruction to a set of two-dimensional reconstructions. The quality of the four-dimensional reconstruction depends entirely on the quality of the individual two-dimensional reconstructions, which is well-understood in terms of projection angles. A lengthy nested scan utilizing this principle has been demonstrated \cite{Jaster-Merz_2024}; however, accurate reconstructions appear to be possible without sampling the entire space of phase advances. Examples include holding one phase advance fixed while the other varies \cite{Wolski_2020} or varying both phases in a single quadrupole scan \cite{Roussel_2023}. These studies suggest that  $\mu_x$, $\mu_y$, and $\mu_x - \mu_y$ should cover a range of values between $0$ and $\pi$.

\begin{figure}
    \centering
    \includegraphics[width=0.65\columnwidth]{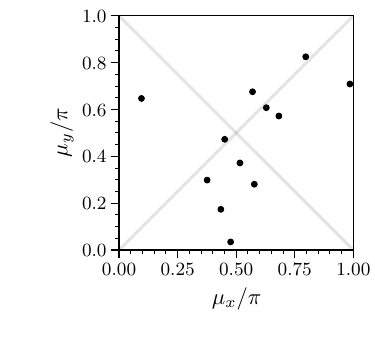}
    \caption{Phase advances from the reconstruction point to each wire scanner. Three sets of optics and four wire scanners per measurement give twelve data points.}
    \label{fig:phase-advances}
\end{figure}

The constraints in the SNS limit our control of the phase advances at each wire scanner. In \cite{Hoover_2022}, we found that the nominal optics produce little variation in the difference $\mu_x - \mu_y$, leading to an ill-conditioned linear system when fitting the covariance matrix. Because of the shared quadrupole power supplies (Fig.~\ref{fig:ws}), there are only two control parameters: one scaling parameter for the focusing quadrupoles, and one for the defocusing quadrupoles. These two parameters determine the phase advances from the reconstruction point to the last wire scanner (WS24). A constrained optimization (ensuring a small downstream beam size and a fixed beam size at the target) led to a second set of optics with a better-conditioned transfer matrix. In the second set of optics, the phase advances $\mu_x$ and $\mu_y$ at WS24 were shifted by 45 degrees in opposite directions relative to their nominal values. In the present study, we added a third set of optics with different phase advances at WS24. The phase advances at each wire scanner over the three sets of optics are plotted in Fig.~\ref{fig:phase-advances}. 

\subsection{Beam accumulation}

The SNS employs phase space painting to shape the transverse particle distribution. Particles are injected at an offset in both planes and without transverse momentum ($x, y > 0$, $x' = y' = 0$). Kicker magnets increase $x$ and $y$ during accumulation, leading to an approximate doughnut distribution in $x$-$x'$ and $y$-$y'$. (There is no coupling in the ring.) Collective effects tend to fill in the doughnuts and generate an approximately uniform $x$-$y$ density within a rectangular boundary. Linear cross-plane correlations are unexpected because they would be washed during injection.

\textit{Eigenpainting} \cite{Holmes_2018} is a proposed painting method which begins by injecting particles on the closed orbit: $x = y = x' = y' = 0$. The injected coordinates are scaled along the real part of an eigenvector of the $4 \times 4$ ring transfer matrix. If a particle is injected onto an eigenvector, its turn-by-turn coordinates jump along an ellipse in the $x$-$y$ plane. Continuous injection at the same amplitude leads to a uniform-density ellipse, with all particles traveling tangent to the ellipse. By varying the ellipse area but fixing its orientation, one may generate a uniform particle density within an elliptical boundary. This is true even with the inclusion of space charge (assuming linear optics and various other approximations). Since particles occupy a single eigenvector, they occupy zero volume in the four-dimensional phase space.

We measured the beam during an initial test of the eigenpainting method. This method required a lower beam energy of 0.8 GeV to boost the effective strength of the injection kicker magnets. We will not describe the details of this experiment here; the important point is that we intended to generate cross-plane correlations in the beam. We measured the beam for the three sets of optics described earlier, generating 36 profiles. We approximated the accelerator transport as a linear transformation of the phase space coordinates and computed the transport matrices using the OpenXAL \cite{Zhukov_2022} online model of the SNS. (Appendix~\ref{app:production} reports a measurement of a production beam; however, we were only able to collect two sets of optics (24 profiles) in this experiment.)

The data and code to reproduce the figures in this paper can be found online \footnote{\url{https://doi.org/10.5281/zenodo.14009840}}.

\section{Results}\label{sec:results}

\begin{figure}
    \centering
    \includegraphics[width=\columnwidth]{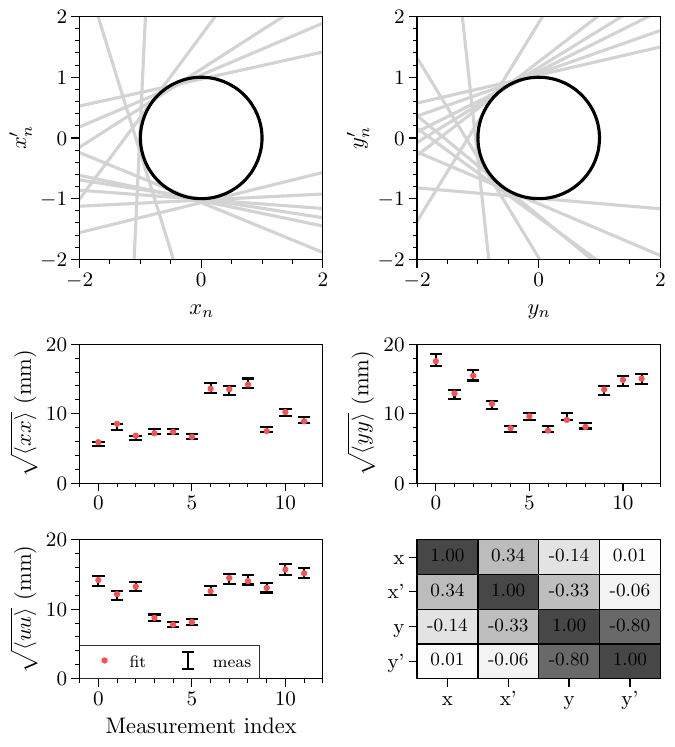}
    \caption{Least-squares fit of the $4 \times 4$ covariance matrix to measured rms beam sizes. Top: vertical lines in $x$-$x'$ and $y$-$y'$ phase space mapped back to the reconstruction point and plotted in normalized coordinates. In these units, the best-fit ellipse is the unit circle. Bottom: measured (black) vs. fit (red) rms beam sizes on each wire, assuming $5\%$ measurement error. Bottom right: best-fit correlation matrix.}
    \label{fig:cov}
\end{figure}

Fig.~\ref{fig:cov} shows the LLSQ fit of the covariance matrix to the measured rms beam sizes.
The fit reproduces the beam sizes on all wires in all scan steps. The tightly fit ellipses over a range of phase advances in the $x$-$y'$ and $y$-$y'$ planes indicate a well-modeled lattice and small measurement errors. LLSQ error propagation gives uncertainties of a few percent for the cross-plane moments. There are some linear cross-plane correlations, particularly between $y$ and $x'$. These correlations vary with the position along the beamline.

\begin{figure*}
    \centering
    \subfloat[][Linear scale]{%
        \includegraphics[width=\columnwidth]{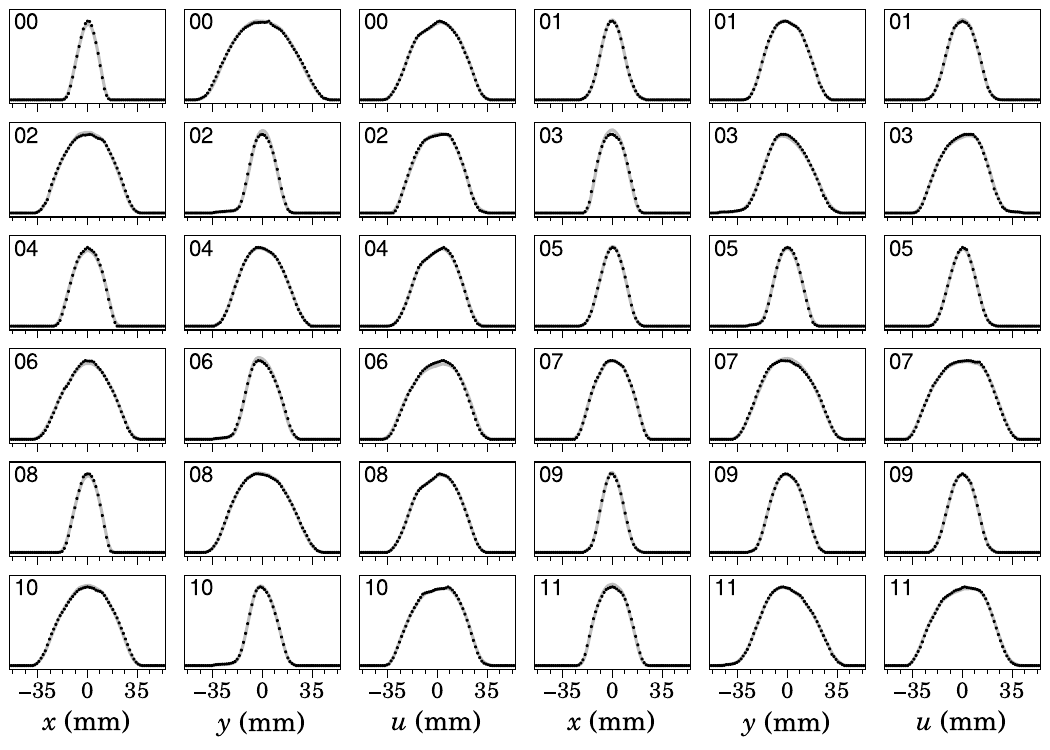}%
    }
    \hfill
    \subfloat[][Log scale]{%
        \includegraphics[width=\columnwidth]{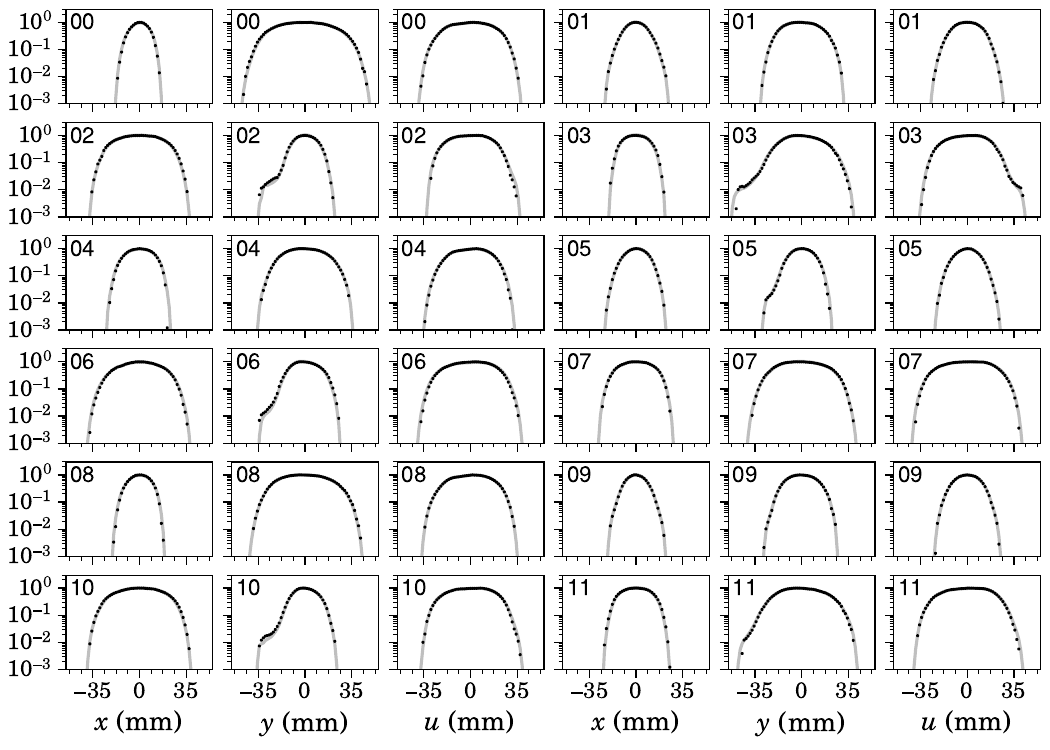}%
    }
    \caption{Measured (black) vs. predicted (gray) density on each wire. The measurement index is printed on the top left of each subplot.}
    \label{fig:profiles}
\end{figure*}

With a Gaussian prior based on the measured covariance matrix, the MENT algorithm converged to the linear-scale profiles in one iteration and the log-scale profiles in another two iterations, terminating at a mean absolute error of $10^{-4}$. Fig.~\ref{fig:profiles} displays the measured and simulated beam profiles. 
Notice the agreement of the profiles down to the $10^{-3}$ level; with improved wire scanner dynamic range, MENT may be able to study halo formation in two- or four-dimensional phase space. We do not know if the asymmetric tails are real or if they are due to background or crosstalk between wires.

\begin{figure}
    \centering
    \includegraphics[width=\columnwidth]{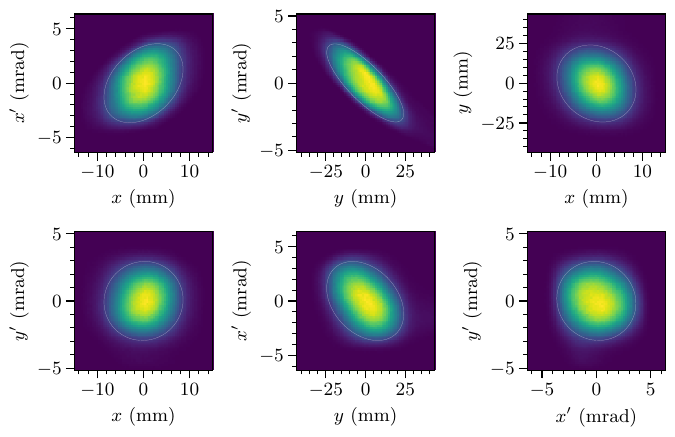}
    \caption{Two-dimensional projections of the reconstructed four-dimensional phase space distribution. RMS ellipses are overlayed.}
    \label{fig:proj2d}
\end{figure}

Fig.~\ref{fig:proj2d} shows the two-dimensional marginal projections of the posterior distribution. 
The posterior distribution is neither perfectly Gaussian nor perfectly uniform, consistent with the one-dimensional projections in Fig.~\ref{fig:profiles}. There is also some asymmetry in the distribution, particularly in the $y$-$y'$ projection; this is again consistent with some of the measured profiles. Space charge and other nonlinear effects are likely responsible for the nonuniformity of the distribution and the blurred cross-plane correlations. A detailed examination of the distribution is beyond the scope of this paper.

\section{Uncertainty Quantification}\label{sec:uncertainty_quantification}

\subsection{Modeling and measurement errors}

One source of reconstruction error is the forward model. Our model represents the accelerator as a $4 \times 4$ transfer matrix:
\begin{equation}
    \vect{M} = 
    \begin{bmatrix}
        \vect{M}_{xx} & \vect{M}_{xy} \\
        \vect{M}_{yx} & \vect{M}_{yy}
    \end{bmatrix},
\end{equation}
with $\vect{M}_{xy} = \vect{M}_{yx} = \vect{0}$. The following are possible errors in this model: (i) The true quadruple coefficients may not be equal to the readback values. We assume these values are accurate because measured twiss parameters typically agree with design values \cite{Hoover_2022}; future experiments could validate the linear model using orbit-difference measurements. (ii) Tilted quadrupoles would generate off-block-diagonal terms in the transfer matrix. We expect skew quadrupole components to be negligible, as they can be measured using BPM data \cite{Pelaia_2008} and would result in a tilted image on the target during neutron production. (iii) An incorrect beam energy would cause a systematic error in the quadrupole focusing strengths. We ignore this source of error because the beam energy is known to within 0.1\% from time-of-flight measurements in the ring  \cite{Wong_2021_lace}. (iv) The quadrupole coefficients depend on the particle energy---each particle sees a slightly different force---but our model ignores energy spread. Strong transverse-longitudinal correlations in the bunch could amplify the error. The nominal energy spread in the SNS is only a few MeV on top of a 1 GeV synchronous particle energy, so we expect that our model is sufficient \footnote{The beam energy in this experiment was 0.8 GeV, leading to a slightly larger fractional energy spread.}. Including energy-dependent focusing would require a six-dimensional reconstruction. (v) The model does not include space charge. The tune shift over one turn in the ring is approximately 3\%, too small to make a difference in the RTBT wire scanner region \cite{Hoover_2022}. (vi) The wire scanner measurements assume the distribution has no pulse-to-pulse variation. Back-to-back measurements show almost no change in the profiles. In the future, an electron scanner \cite{Blokland_2009} could validate this assumption by measuring the full profile on each pulse.

\subsection{Suboptimal views}

Even with a perfect forward model, projections can only place \textit{constraints} on the unknown distribution; they cannot, on their own, identify a unique solution. Ideally, the reconstruction would generate an ensemble of possible solutions. Uncertainty quantification of this sort is challenging and has yet to be incorporated in experimental reconstructions. Instead, to build confidence in the results, many studies have relied on fake-data reconstructions, where the data are generated by a known ground truth distribution, usually from a beam physics simulation. We take a similar approach here.

\begin{figure*}
    \subfloat[][PyORBIT]{%
        \begin{minipage}{0.98\columnwidth}
            \begin{center}
                \includegraphics[width=\columnwidth]{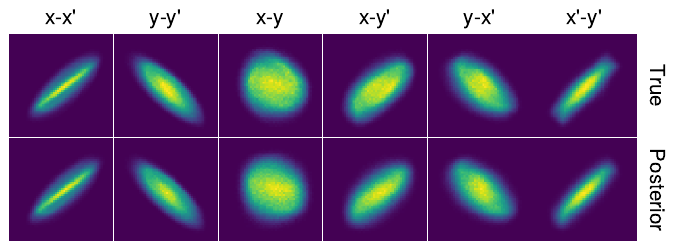}\\%
                \medskip
                \includegraphics[width=\columnwidth]{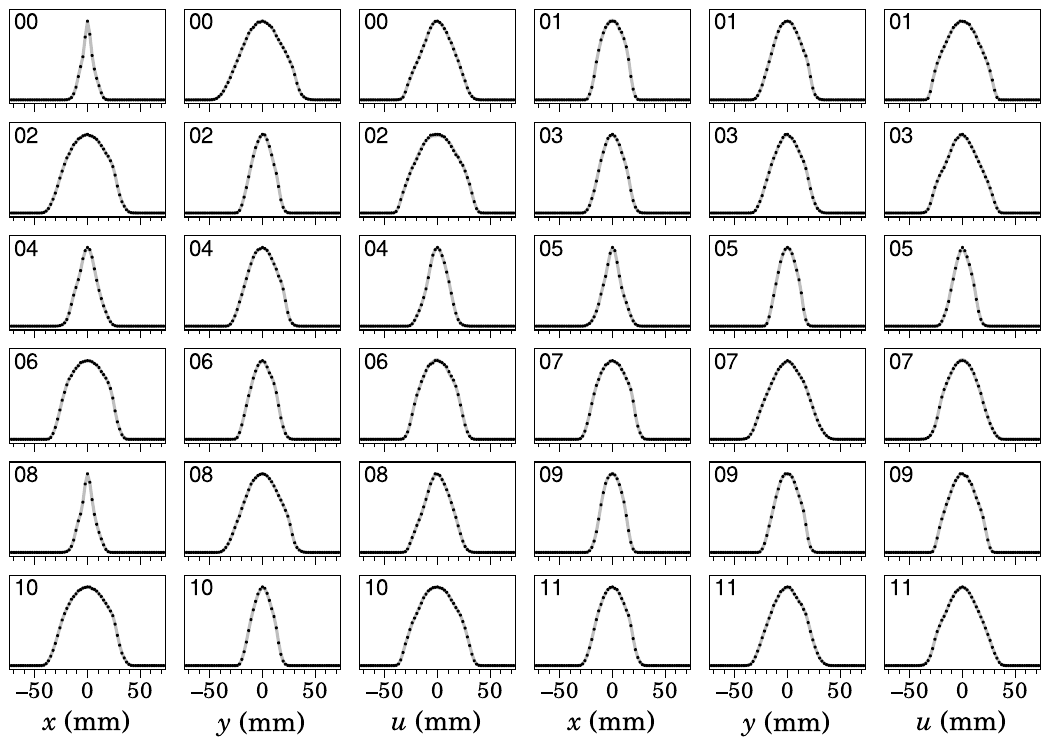}%
            \end{center}
        \end{minipage}
    }%
    \hfill
    \subfloat[][Gaussian Mixture]{%
        \begin{minipage}{0.98\columnwidth}
            \begin{center}
                \includegraphics[width=\columnwidth]{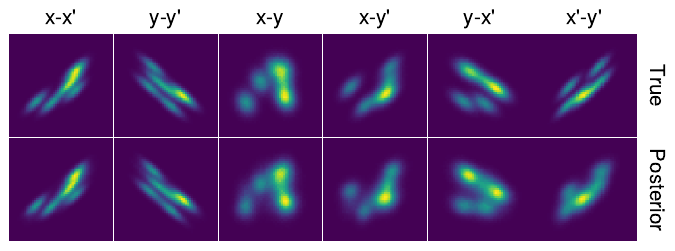}\\%
                \medskip
                \includegraphics[width=\columnwidth]{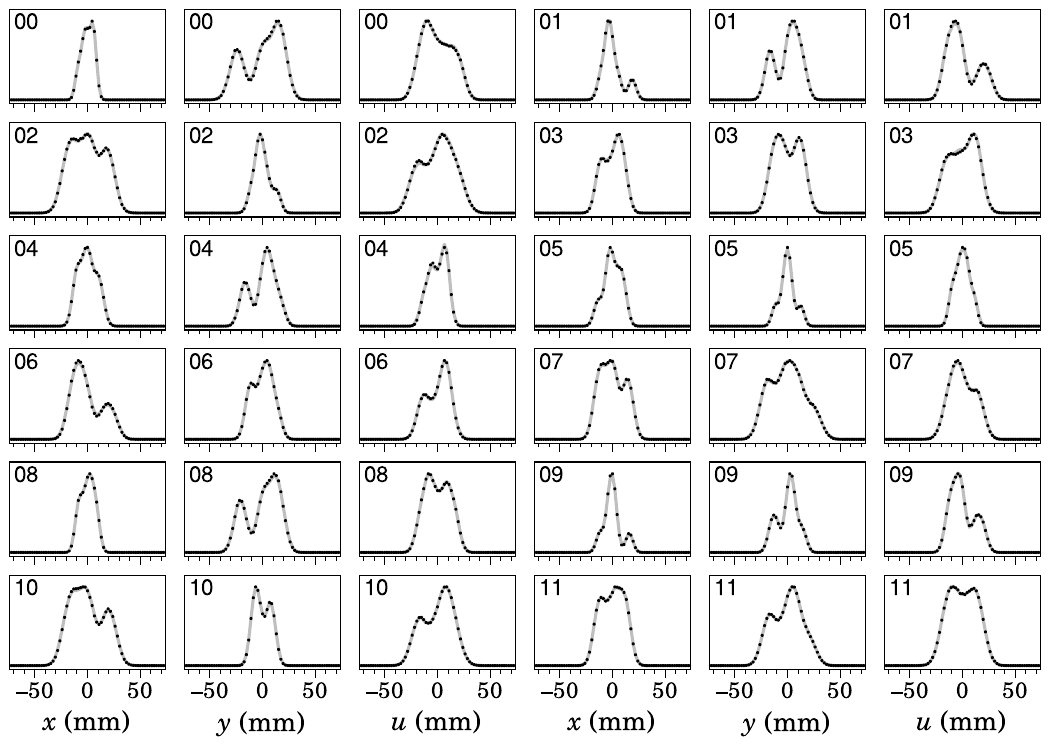}%
            \end{center}
        \end{minipage}
    }%
    \vfill
    \subfloat[][Waterbag]{%
        \begin{minipage}{0.98\columnwidth}
            \begin{center}
                \includegraphics[width=\columnwidth]{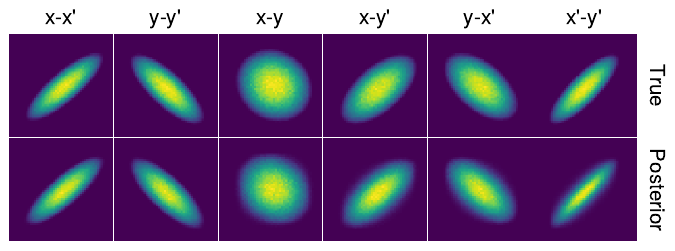}\\%
                \medskip
                \includegraphics[width=\columnwidth]{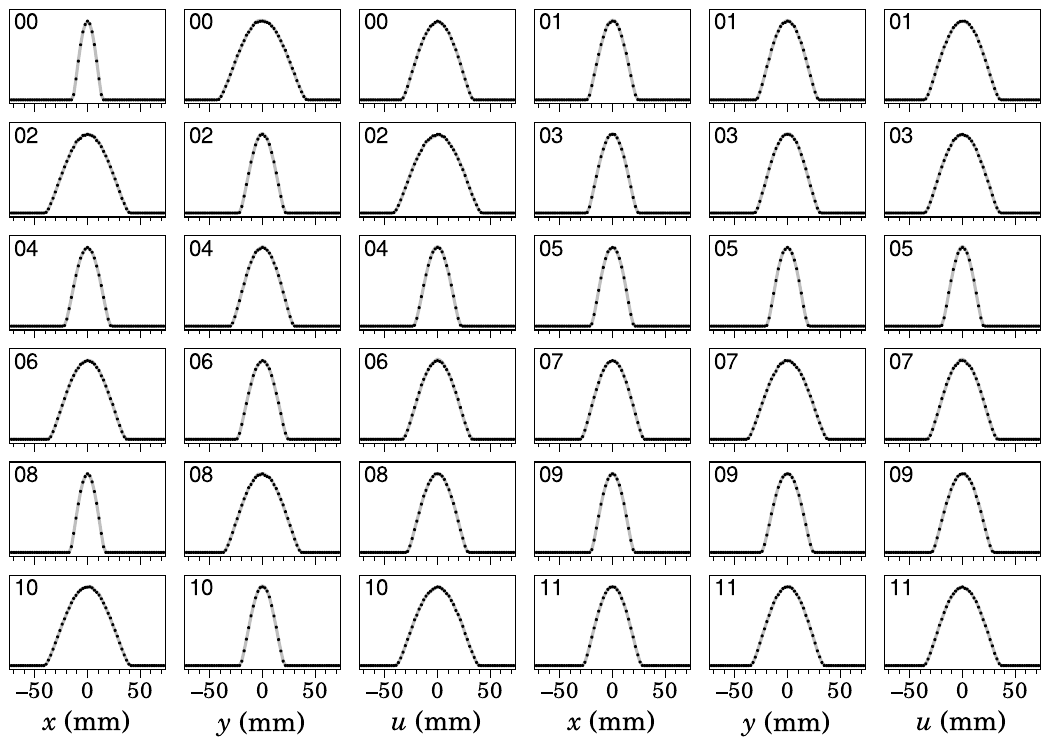}%
            \end{center}
        \end{minipage}
    }%
    \hfill
    \subfloat[][Hollow]{%
        \begin{minipage}{0.98\columnwidth}
            \begin{center}
                \includegraphics[width=\columnwidth]{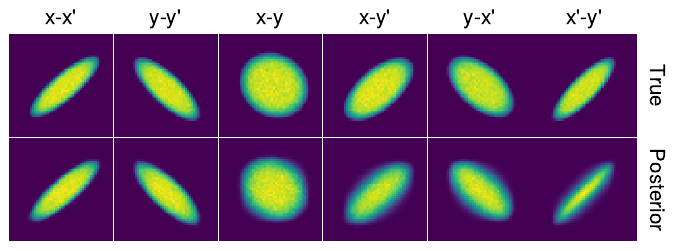}\\%
                \medskip
                \includegraphics[width=\columnwidth]{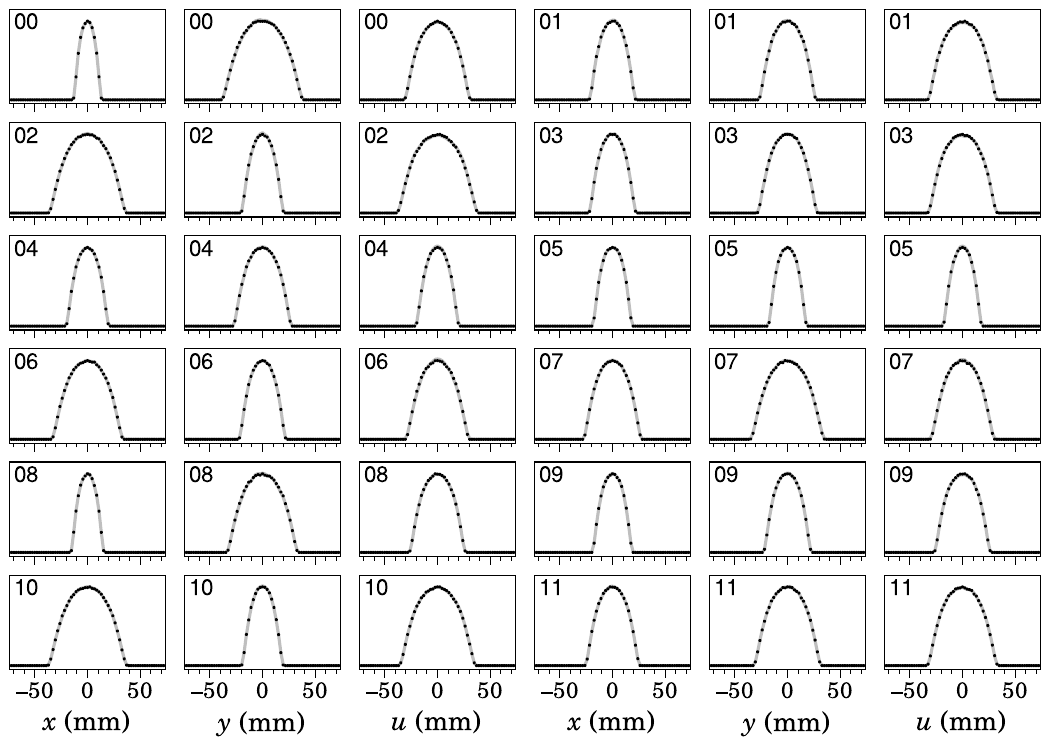}%
            \end{center}
        \end{minipage}
    }%
    \caption{Fake-data reconstructions. Each subplot (a-d) corresponds to a different initial distribution. Each subplot shows the two-dimensional projections of the posterior and ground truth distributions, followed by the predicted vs. true profiles at the measurement locations. True profiles are plotted as black dots; predicted profiles are plotted as gray lines.}
    \label{fig:sim}
\end{figure*}

We selected four distributions: (i) the result of a PyORBIT \cite{Shishlo_2015} simulation of a similar injection scheme (reported in \cite{Holmes_2018}); (ii) a uniform distribution inside the unit sphere (Waterbag): 
\begin{equation} 
    f(\mathbf{v}) \propto \Theta(1 - |\mathbf{v}|),
\end{equation}
where $\Theta$ is the unit step function; (iii) a hollow distribution (Hollow):
\begin{equation} 
    f(\mathbf{v}) \propto \Theta(1 - |\mathbf{v}|) |\mathbf{v}|^\gamma,
\end{equation}
with $\gamma = 4$; and (iv) a Gaussian mixture distribution (GM), which is the superposition of seven Gaussian distributions with random means and variances: 
\begin{equation} 
    f(\mathbf{z}) \propto \sum_{i} \mathcal{N}(\mathbf{\mu}_i, \mathbf{\sigma}_i).
\end{equation}
Samples from each distribution were linearly transformed to match the measured covariance matrix. Fig.~\ref{fig:sim} shows the simulated reconstruction results. 

The PyORBIT distribution represents a best-case scenario for the injection method: its density is somewhat uniform within an elliptical boundary in the $x$-$y$ plane, and one of its eigenemittances is much smaller than the other. These features are usually clear from two-dimensional views. In Fig.~\ref{fig:sim}, the two-dimensional projections of the posterior distribution generally agree with the ground truth. Projections of the GM distribution show that this set of optics should also unveil more complex two-dimensional patterns. The GM projections indicate a greater uncertainty in the cross-plane projections, particularly $x'$-$y'$, where the posterior does not break into disconnected modes. Nonetheless, these two examples raise our confidence in the two-dimensional projections of the posterior distribution.

We observe similar agreement for the Waterbag and Hollow distributions, although the Hollow distribution's projections are not as uniform as they should be. One must be cautious when comparing distributions using only low-dimensional views. Projections are less informative when they average over many parts of the distribution. Low-dimensional projections average over many dimensions and can hide internal structure such as hollowing in the beam core \cite{Cathey_2018}. It follows that hollow distributions are more difficult to reconstruct than peaked distributions. This was partially explored in \cite{Hoover_2024} for the six-dimensional case. 

\begin{figure}
    \centering
    \includegraphics[width=\columnwidth]{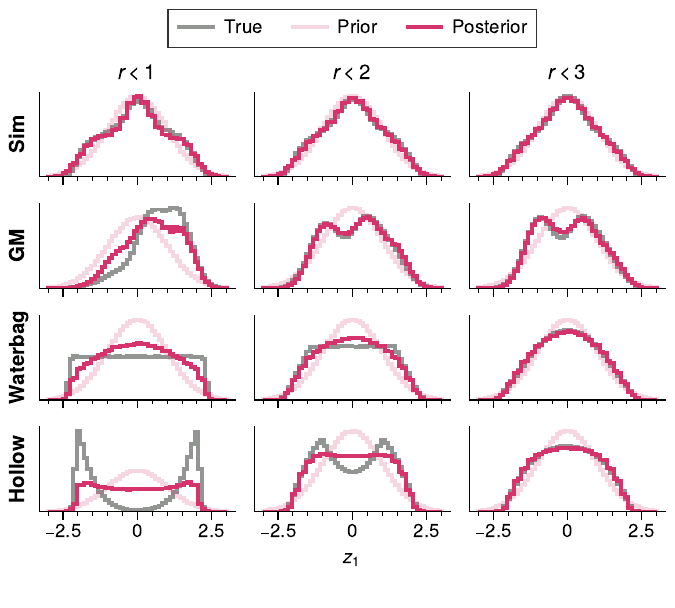}
    \caption{Sliced views of the simulated reconstructions in Fig.~\ref{fig:sim}. The four-dimensional distributions are normalized to remove linear correlations and scaled to unit variance. For normalized coordinates $\vect{z} = (z_1, z_2, z_3, z_4)$, the distribution $f(z_1, z_2, z_3, z_4)$ is sliced to select particles in a ball of radius $r = (z_2^2 + z_3^2 + z_4^2)^{1/2}$. The figure shows the $z_1$ coordinates within this slice. The slices in the right column select almost all particles, while the slices in the left column select particles near the core in the $z_2$-$z_3$-$z_4$ plane. In each subplot, the true distribution and the posterior distribution generate the same projections on the measurement axes.}
    \label{fig:sim_spherical_slices}
\end{figure}

Fig.~\ref{fig:sim_spherical_slices} reveals errors in the internal structure of the Hollow distribution and, to some extent, the Waterbag distribution.
The figure shows the particle distribution along one dimension as the other three dimensions are sliced. The slice selects particles within a ball of radius $r$ in the unplotted three-dimensional space. When $r$ is large, the slice selects all particles, but when $r$ is small, it selects particles near the core. In the bottom left subplot of Fig.~\ref{fig:sim_spherical_slices}, we see that the posterior distribution has failed to excavate the core. MENT flattens the Gaussian prior until the constraints are satisfied, but no more.

We conclude that the measurement optics are informative: they should reveal most two-dimensional features. For our purposes at the SNS, this is sufficient to predict the beam density on the target and to benchmark simulation models with improved precision. Additional measurements may be necessary to capture the full four-dimensional structure. It may be possible to tighten the constraints using the same number of measurements, but the shared quadrupole power supplies and constraints on the beam size do not leave much room for exploration. The measurement optics were chosen to reduce the uncertainty in the covariance matrix; until a different criterion emerges, this seems to be an effective strategy to reduce uncertainty in the tomographic reconstruction. These results highlight the need for robust uncertainty quantification techniques.

\section{Conclusion}\label{sec:conclusion}

We have used one-dimensional measurements to reconstruct the four-dimensional phase space distribution in the SNS accelerator. We performed the reconstruction using the MENT algorithm with a Gaussian prior defined by the best-fit covariance matrix. Fake-data reconstructions with different initial distributions indicate that the measurements are informative and should reproduce most two-dimensional features. Additional constraints may be needed to fully capture higher-dimensional features such as four-dimensional hollowing. We conclude that four-dimensional tomography can be useful when only one-dimensional measurements are available.

A possible use of four-dimensional tomography in the SNS is to predict the beam density on the spallation target \cite{Blokland_2014} or the planned Second Target Station \cite{Galambos_2015}. Another possible use is to benchmark simulation codes: extracting and measuring the beam on different turns would provide an estimate of the four-dimensional distribution as a function of time. Finally, an electron scanner \cite{Blokland_2009} could provide turn-by-turn profiles in the ring, and it may be possible to infer the distribution from these profiles if the beam is mismatched. Because of the large volume of data from the electron scanner and the importance of collective effects in the ring, surrogate-model-based reconstruction algorithms may be the most effective option \cite{Scheinker_2021}.

\section{Acknowledgments}
I would like to thank Wim Blockland (ORNL) for troubleshooting the RTBT wire scanners, the SNS accelerator operators for enabling these studies, and Andrei Shishlo (ORNL) for carefully reading this manuscript.

This manuscript has been authored by UT Battelle, LLC under Contract No. DE-AC05-00OR22725 with the U.S. Department of Energy. The United States Government retains and the publisher, by accepting the article for publication, acknowledges that the United States Government retains a non-exclusive, paid-up, irrevocable, world-wide license to publish or reproduce the published form of this manuscript, or allow others to do so, for United States Government purposes. The Department of Energy will provide public access to these results of federally sponsored research in accordance with the DOE Public Access Plan (http://energy.gov/downloads/doe-public-access-plan).

\appendix

\section{Measurement of a production beam}\label{app:production}

\begin{figure}
    \includegraphics[width=\columnwidth]{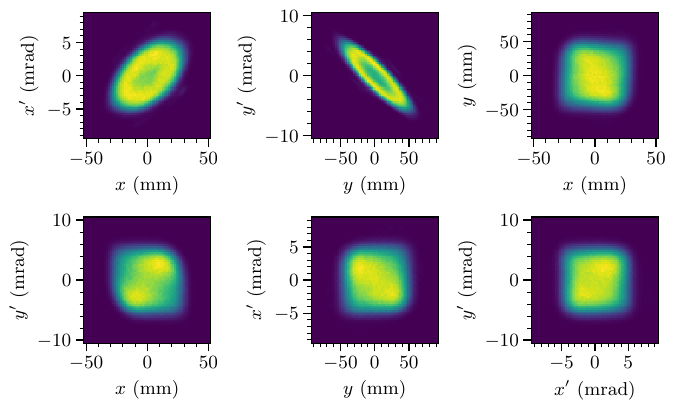}%
    \caption{Reconstructed production beam from two sets of optics (24 profiles).}
    \label{fig:exp02}
\end{figure}

This appendix reports the measurement of a production beam in the SNS. By ``production beam'', we mean that the accelerator was left in its default state, tuned for losses during neutron production at 1.4 MW beam power. Fig.~\ref{fig:exp02} shows the reconstructed distribution. 

Due to time constraints, we only used two sets of optics in this experiment. Two sets of optics are sufficient for the covariance matrix fit, but we have not studied the effect on the distribution. Still, the projections are largely in line with expectations. The doughnut-shaped $x$-$x'$ and $y$-$y'$ projections follow from the painting method, which injects particles with an initial offset. Collective effects tend to fill in the hollow core. Linear cross-plane correlations should be negligible, and the cross-plane projections, such as $x$-$y$, should be somewhat uniform within a rectangular boundary. Both features are present in the reconstruction.

We leave uncertainty quantification as future work.

\bibliography{main}

\end{document}